\documentclass{iau} 
\usepackage{graphicx}

\title[~~Inclusive education and research] 
{Inclusive education and research through African Network of Women in Astronomy and STEM for GIRLS in Ethiopia initiatives}

\author[Mirjana Povi\'c et al.]   
{Mirjana Povi\'c$^{1,2}$, Vanessa McBride$^3$, Priscilla Muheki$^4$, Carolina \"Odman-Govender$^5$, Somaya Saad${^6}$, Nana Ama Brown Klutse${^7}$, Aster Tsegaye${^8}$, Tigist Getachew${^9}$, Melody Kelemu${^{10}}$, Hanna Kibret$^1$, Jerusalem Tamirat$^1$, Deborah Telahun-Teka${^9}$, Beza Tesfaye$^{11}$, and Feven Tigistu-Sahle${^9}$}

\affiliation{
$^1$Ethiopian Space Science and Technology Institute, Ethiopia, email: {\tt mpovic@iaa.es}. $^2$Instituto de Astrof\'isica de Andaluc\'ia (CSIC), Spain. $^3$Office of Astronomy for Development-IAU, South Africa. $^4$Mbarara University of Science and Technology, Uganda. $^5$Inter-University Institute for Data Intensive Astronomy, University of the Western Cape, South Africa. $^6$National Research Institute of Astronomy and Geophysics, Egypt. $^7$University of Ghana, Ghana. $^8$Addis Ababa University, Ethiopia. $^9$Ethiopian Biotechnology Institute, Ethiopia. $^{10}$International Institute for Primary Health Care, Ethiopia. $^{11}$Ethiopian Space Science Society, Ethiopia. \\[\affilskip]
	}

\pubyear{2021}
\volume{367}  
\jname{Education and Heritage in the era of Big Data in Astronomy}
\editors{R.M. Ros, B. Garcia, S. Gullberg, J. Moldon \& P. Rojo, eds.}
\begin{document}

\maketitle

\begin{abstract}
The African Network of Women in Astronomy and STEM for GIRLS in Ethiopia initiatives have been established with aim to strengthen the participation of girls and women in astronomy and science in Africa and Ethiopia. We will not be able to achieve the UN Sustainable Development Goals without full participation of women and girls in all aspects of our society and without giving in future the same opportunity to all children to access education independently on their socio-economical status. In this paper both initiatives are briefly introduced. 

\end{abstract}

\firstsection 

\vspace{-4.5mm}

\section{African Network of Women in Astronomy- AfNWA}
Considering the latest report of the UNESCO and UN-WOMEN, the number of female researchers in the world (both part- and full-time) is on average $<$\,30\% (\cite[UNESCO, 2019]{unesco2019}). For most of countries this number becomes even lower when STEM (Science, Technology, Engineering and Mathematics) fields are considered. Therefore, globally we are facing significant gender gap in science. In Africa, most of countries have a number of female scientists below 25\%. 
Many factors may be responsible for the low number of female scientists (e.g., poverty and lack of access to education, social constraints, cultural biases and beliefs, lack of female mentors and role models, etc.), but the final result is that these difficulties mean we are losing huge potential that could benefit our society. We will never be able to reach the UN Sustainable Development Goals (SDGs) without giving our best in empowering girls and women who make $\sim$\,50\% of world population. Astronomy and space sciences are currently experiencing significant growth in Africa (\cite[Povi\'c et al., 2018]{povic2018}). For the benefit of all society we would like to guarantee future participation of girls and women at all levels in astronomy and science developments in Africa.

The African Network of Women in Astronomy (AfNWA)\footnote{https://www.africanastronomicalsociety.org/afnwa/} is an initiative established in September 2020 under the African Astronomical Society (AfAS)\footnote{https://www.africanastronomicalsociety.org/} that aims to connect women working in astronomy and related fields in Africa. Our main objectives are improving the status of women in science in Africa, and using astronomy to inspire more girls to do STEM. These objectives are planned to be achieved through different activities such as: the creation of AfNWA network, organisation of needed courses and trainings for improving research and leadership skills of women in astronomy in Africa, giving visibility to women in astronomy and related fields (through yearly published reports, newsletters, website, public talks, public communications, given awards, etc.), organisation of outreach activities given by women astronomers, understanding of the main factors responsible for the lack of women in astronomy and science in different African countries, and development of efficient methods of retaining women in astronomy.

Since AfNWA objectives are there for the benefit of the whole society, we strongly encourage participation and support of the whole population in our activities, both women and men.

\vspace{-5mm}

\section{STEM for GIRLS in Ethiopia}

In the last Ethiopian Growth and Transformation Plan (GTP) II (2016 - 2020) it has been raised that
women in Ethiopia face multiple challenges, including illiteracy and inequality in education,
unequal division of labour, unequal power relationships, and limited participation in leadership and decision-making (\cite[GTP II, 2016]{gtp2016}). In addition, participation of young girls in STEM and women in science is still far from reaching gender balance due to different social and economical challenges. STEM for GIRLS in Ethiopia initiative was established in 2019 in collaboration with the Society of Ethiopian Women in Science and Technology (SEWiST) with aim to help to achieve the goals of the GTP II and to improve in future the participation of women and girls in STEM. It is based on creating a strong connection between SEWiST members and the rest of society, in particular grade 9 and 10 girls and their teachers, and use SEWiST's strength and long experience to promote women in science and technology as role models. Beside inspiring and encouraging more girls to do STEM, we also aim in understanding what are the main factors responsible for the lack of girls and women in STEM in Ethiopia and how can we improve it in future. 
To reach proposed objectives, we organised different activities with girls along 2019 and beginning of 2020. Almost all activities were carried out in Addis Ababa, in average once per month. We organised interactions between girls and women scientists, so that women can share their professional experiences together with their life story on how they became scientists and therefore serve as role models, and that girls can share their life stories as well, their future plans, challenges they are facing, etc. Up to now the interaction has been made with almost 1000 girls, with many activities being focused on astronomy. We also developed a questionnaire that girls can feel voluntarily, for understanding better the reality and main factors behind the lack of girls in STEM, seeing clearly the tendency of girls to continue with care-type professions (up to 70\%). In addition, in 2019 we organised the very first workshop for teachers to discuss why it is important to bring more girls into STEM. Our activities were interrupted during 2020 due to the COVID-19 pandemic and closure of schools in Ethiopia, but we are now re-starting the activities. 

\vspace{0.1cm}

\textbf  {\small{Acknowledgements}}\\
\small{AfNWA and STEM for GIRLS in Ethiopia have been supported by the Nature Research and Est\'ee Lauder through the MP 2018 Inspiring Science Award. Support of the South African DSI to AfAS-AfNWA is also highly acknowledged.}

\end{document}